\documentclass[a4paper]{jpconf}
\usepackage{graphicx}
\begin{document}
\title[c]{A model for projectile fragmentation}
\author{\large G. Chaudhuri$^{1*}$, S. Mallik$^1$ and S. Das Gupta$^2$}
\address{$^1$Theoretical Physics Division, Variable Energy Cyclotron Centre, 1/AF Bidhan Nagar, Kolkata 700064, India}
\address{$^2$Physics Department, McGill University, Montr{\'e}al, Canada H3A 2T8}
\ead{$^{*}$gargi@vecc.gov.in}
\begin{abstract}
A model for projectile fragmentation is developed whose origin can be traced back to the Bevalac era.  The model positions itself between the phenomenological EPAX parametrization and transport models like ``Heavy Ion Phase Space Exploration'' (HIPSE) model and antisymmetrised molecular dynamics (AMD) model. A very simple impact parameter dependence of input temperature is incorporated in the model which helps to analyze the more peripheral collisions. The model is applied to calculate the charge, isotopic distributions, average number of intermediate mass fragments and the average size of largest cluster at different $Z_{bound}$  of different projectile fragmentation reactions at different energies.
\end{abstract}

\section{Introduction}

Projectile fragmentation is an important technique for studying the reaction mechanisms in heavy ion collisions at
intermediate and high energies. In heavy ion collisions, if the beam energy is high enough, the participant-spectator scenario can be envisaged. The participant zone is highly excited whereas the projectile like fragment (PLF) with rapidity close to that of the projectile rapidity and the target like fragment (TLF)  with rapidity near zero are mildly excited. The PLF has been studied experimentally, this being one of the tools for production and identification of exotic nuclei. A projectile fragmentation model is developed \cite{Mallik2, Mallik3} which involves concepts of heavy ion reaction plus the well known statistical model of multifragmentation (Canonical Thermodynamical Model) and evaporation.  This model is computationally much less intensive  than  heavy ion phase-space exploration (HIPSE) model \cite{Lacroix} and antisymmetrized molecular dynamics (AMD) \cite{Ono} which are based on transport calculation and less phenomenological than EPAX \cite{Summerer} which is based on the empirical parametrization of fragmentation cross sections . An impact parameter dependent temperature profile has been developed in order to better account for the results at different $Z_{bound}$ ranges and also to confront with data from different projectile fragmentation reactions at different energies. The model is in general applicable and implementable above 100 MeV/nucleon.\\
The organization of the article is as follows. In Sec.2 we describe the theoretical formulation of the model where as the impact parameter dependence of temperature is explained in Sec.3. Sec.4 contains the results obtained from theoretical calculation and comparison with experimental data of different projectile fragmentation reactions with different projectile-target combinations and varying projectile energies. Finally summary and conclusions are presented in Sec.5.

\section{Formulation of Model}

The model for projectile fragmentation reaction consists of three stages: (i) abrasion, (ii) multifragmentation and (iii) evaporation. In heavy ion collision, if the beam energy is high enough, then in the abrasion stage at a particular impact parameter three different regions are formed: (i) projectile spectator or projectile like fragment (PLF) moving in the lab with roughly the velocity of the beam, (ii) participant which suffer direct violent collisions and (iii) target spectator or target like fragment (TLF) which have low velocities in the lab. Here we are interested in  the fragmentation of the PLF. The number of neutrons and protons in the projectile spectator at different impact parameters are determined from abrasion stage. Then the break up of each abraded projectile spectator is separately calculated by using canonical thermodynamical model (CTM) \cite{Das}. Finally, the decay of excited fragments are calculated by evaporation model \cite{Mallik1} based on Weisskopf's formalism. We describe the details of the three different stages below.

\subsection{Abrasion}

In abrasion stage we assume the beam energy is high enough so that using straight-line geometry, PLF, TLF and participant can be classified.  We then calculate the volume of the projectile that goes into the participant region (eqs. A.4.4 and A.4.5 of \cite{Dasgupta1}).  What remains in the PLF is $V$.  This is a function of $b$, the impact parameter. If the original volume of the projectile is $V_0$, the original number of neutrons is $N_0$
and the original number of protons is $Z_0$ then the average number of neutrons in the PLF is $<N_s(b)>=(V(b)/V_0)N_0$ and the average number of protons is $<Z_s(b)>=(V(b)/V_0)Z_0$.  These will usually be non-integers. Since in any event only integral numbers for neutrons and protons can materialise in the PLF, we have to guess what is the distribution of $N_s,Z_s$ which produces these average values. For this purpose minimal distribution model is used.  Let $<N_s(b)>=N_s^{min}(b)+\alpha$ where $\alpha$ is less than 1.  We can also define $N_s^{max}(b)=N_s^{min}(b)+1$.  We assume that $P_N(b)$ is zero
unless $N_s(b)$ is $N_s^{min}(b)$ or $N_s^{max}(b)$.  The distribution is narrow. We then get $P(N_s^{max}(b))=\alpha$ and $P(N_s^{min}(b))=1-\alpha$. From $<Z_s>$ we can similarly define $P_{Z_s(b)}$.  Together now we write $P_{N_{s},Z_{s}}(b)=P_{N_{s}}(b)P_{Z_{s}}(b)$.

The abrasion cross-section when there are $N_s$ neutrons and $Z_s$ protons in the PLF is labelled by $\sigma_{a,N_{s},Z_{s}}$:
\begin{equation}
\sigma_{a,N_{s},Z_{s}}=\int 2\pi bdbP_{N_{s},Z_{s}}(b)
\end{equation}
where the suffix $a$ denotes abrasion. The limits of integration in eq.(1) are $b_{min}$ and $b_{max}=R_{target}+ R_{projectile}$. For $b_{min}$ we have either 0 (if the projectile is larger than the target) or $R_{target}-R_{projectile}$ (if the target is larger than the projectile, in this case at lower value of $b$ there is no PLF left). Actually there is an extra parameter that needs to be specified.  The complete labeling is $\sigma_{a,N_s,Z_s,T}$ if we assume that irrespective of the value of $b$, the PLF has a temperature $T$.  Here we have broadened this to the more general case where the temperature is dependent on the impact parameter $b$.  In evaluating eq.(1) we replace integration by a sum.  We divide the interval $b_{min}$ to $b_{max}$ into small segments of length $\Delta b$.  Let the mid-point of the $i$-th bin be $<b_i>$ and the temperature for collision at $<b_i>$ be $T_i$.  Then
\begin{equation}
\sigma_{a,N_s,Z_s}=\sum_i\sigma_{a,N_s,Z_s,T_i}
\end{equation}
where
\begin{equation}
\sigma_{a,N_s,Z_s,T_i}=2\pi<b_i>\Delta bP_{N_S,Z_s}(<b_i>)
\end{equation}
PLF's with the same $N_s,Z_s$ but different $T_i$'s are treated independently for further calculations.
\subsection{Multifragmentation}

The abraded system of $N_s$ neutrons and $Z_s$ protons created at impact parameter b will have an excitation which we characterize by a temperature $T$. The impact parameter dependence of the temperature profile is obtained from different projectile fragmentation reactions with different projectile target combinations. The details of it is described in the next section. The abraded system with $N_s,Z_s$ and a temperature $T$ will break up into many composites and nucleons.  We use the canonical thermodynamic model (CTM) \cite{Das} to calculate this break up. Assume that the system with $N_s$ neutrons and $Z_s$
protons at temperature $T$, has expanded to a higher than normal volume and the partitioning into different composites can be calculated according to the rules of equilibrium statistical mechanics.  In a canonical model, the partitioning is done such that all partitions have the correct $N_s, Z_s$.
The canonical partition function is given by
\begin{eqnarray}
Q_{N_s,Z_s}=\sum\prod \frac{\omega_{N,Z}^{n_{N,Z}}}{n_{N,Z}!}
\end{eqnarray}
Here the sum is over all possible channels of break-up (the number of such channels is enormous) which satisfy $N_s=\sum N\times n_{N,Z}$ and $Z_s=\sum Z\times n_{N,Z}$; $\omega_{N,Z}$ is the partition function of one composite with neutron number $N$ and proton number $Z$ respectively and $n_{N,Z}$ is the number of this composite in the given channel. The one-body partition function $\omega_{N,Z}$ is a product of two parts: one arising from the translational motion of the composite and another from the intrinsic partition function of the composite:
\begin{eqnarray}
\omega_{N,Z}=\frac{V}{h^3}(2\pi mT)^{3/2}A^{3/2}\times z_{N,Z}(int)
\end{eqnarray}
Here $A=N+Z$ is the mass number of the composite and $V$ is the volume available for translational motion; $V$ will be less than $V_f$, the volume to which the system has expanded at break up (freeze-out volume). We use $V = V_f - V_0$ , where $V_0$ is the normal volume of nucleus with $Z_s$ protons and $N_s$ neutrons.  In the projectile fragmentation model we have used a fairly typical value $V_f=3V_0$.

The probability of a given channel $P(\vec n_{N,Z})\equiv P(n_{0,1},n_{1,0},n_{1,1}......n_{I,J}.......)$ is given by
\begin{eqnarray}
P(\vec n_{N,Z})=\frac{1}{Q_{N_s,Z_s}}\prod\frac{\omega_{N,Z}^{n_{N,Z}}}{n_{N,Z}!}
\end{eqnarray}
The average number of composites with $N$ neutrons and $Z$ protons is seen easily from the above equation to be
\begin{eqnarray}
n_{N,Z}=\omega_{N,Z}\frac{Q_{N_s-N,Z_s-Z}}{Q_{N_s,Z_s}}
\end{eqnarray}
There are two constraints: $N_s=\sum N\times n_{N,Z}$ and $Z_s=\sum Z\times n_{N,Z}$. Substituting eq.(7) in these two constraint conditions, two recursion relations \cite{Chase} can be obtained. Any one recursion relation can be used for calculating $Q_{N_s,Z_s}$. For example
\begin{eqnarray}
Q_{N_s,Z_s}=\frac{1}{N_s}\sum_{N,Z}N\omega_{N,Z}Q_{N_s-N,Z_s-Z}
\end{eqnarray}

We list now the properties of the composites used in this work.  The proton and the neutron are fundamental building blocks thus $z_{1,0}(int)=z_{0,1}(int)=2$
where 2 takes care of the spin degeneracy.  For deuteron, triton, $^3$He and $^4$He we use $z_{N,Z}(int)=(2s_{N,Z}+1)\exp(-\beta E_{N,Z}(gr))$ where $\beta=1/T, E_{N,Z}(gr)$ is the ground state energy of the composite and $(2s_{N,Z}+1)$ is the experimental spin degeneracy of the ground state.  Excited states for these very low mass nuclei are not included. For mass number $A=5$ and greater we use the liquid-drop formula.  For nuclei in isolation, this reads ($A=N+Z$)
\begin{eqnarray}
z_{N,Z}(int)&=&\exp\frac{1}{T}[W_0A-\sigma(T)A^{2/3}-a^{*}_c\frac{Z^2}{A^{1/3}}-C_{sym}\frac{(N-Z)^2}{A}+\frac{T^2A}{\epsilon_0}]
\end{eqnarray}
The expression includes the volume energy [$W_0=15.8$ MeV], the temperature dependent surface energy
[$\sigma(T)=\sigma_{0}\{(T_{c}^2-T^2)/(T_{c}^2+T^2)\}^{5/4}$ with $\sigma_{0}=18.0$ MeV and $T_{c}=18.0$ MeV], the Coulomb energy with Wigner-Seitz approximation \cite{Bondorf} [$a^{*}_c=a_{c}\{1-(V_{0}/V_{f})^{1/3}\}$ with $a_{c}=0.72$ MeV] and the symmetry energy ($C_{sym}=23.5$ MeV).  The term $\frac{T^2A}{\epsilon_0}$ ($\epsilon_{0}=16.0$ MeV) represents contribution from excited states since the composites are at a non-zero temperature.

We also have to state which nuclei are included in computing $Q_{N_s,Z_s}$ (eq.(8)). For $N,Z$, (the neutron and the proton number) we include a ridge along the line of stability.  The liquid-drop formula above also gives neutron and proton drip lines and the results shown here include all nuclei within the boundaries.

We repeat the entire break up calculation for each projectile spectator created after abrasion stage with different temperatures at different impact parameters.
Let, $n_{N.Z}^{N_s,Z_s,T_i}$ be the average number of fragment having $N$ neutron and $Z$ proton created after the multifragmentation of a projectile spectator ($N_s,Z_s$) at temperature $T_i$, then cross-section after multifragmentation stage can be expressed as
\begin{equation}
\sigma_{m,N,Z,T_i}=\sum_{N_s,Z_s}n_{N,Z}^{N_s,Z_s,T_i}\sigma_{a,N_s,Z_s,T_i}
\end{equation}
\subsection{Evaporation}
The excited fragments produced after multifragmentation decay to their stable ground states. Its can $\gamma$-decay to shed its energy but may also decay by light particle emission to lower mass nuclei.  We include emissions of $n,p,d,t,^3$He and $^4$He. Particle decay widths are
obtained using the Weisskopf's evaporation theory \cite{Weisskopf}. Fission is also included as a de-excitation channel though for the nuclei of mass $<$ 100 its role will be quite insignificant.
According to Weisskopf's conventional evaporation theory, the partial decay width for emission of a light particle of type $\nu$ is given by
\begin{equation}
 \Gamma_{\nu} = \frac {gm\sigma_0}{\pi^{2}\hbar^{2}} \frac {(E^*-E_0-V_\nu)}{a_R} \exp({2 \sqrt{a_R(E^*-E_0-V_\nu)}-2\sqrt{a_PE^*}})
\end{equation}
For the emission of giant dipole $\gamma$-quanta we take the formula given by Lynn\cite{Lynn}
\begin{equation}
\Gamma_{\gamma}={3 \over \rho_{P}(E^{*})}\int_{0}^{E^*}d\varepsilon\rho_{R}(E^*-\varepsilon)f(\varepsilon)\label{4l}
\end{equation}

For the fission  width we have used the simplified formula of Bohr-Wheeler given by
\begin{equation}
\Gamma_f = \frac{T_P}{2\pi}\exp{(-B_f/T_P)}
\end{equation}
Details of the each term of the above three equations and implementation of the evaporation model can be found \cite{Mallik1}; here we give the essentials necessary to follow the present work.
Once the emission widths ($\Gamma$'s) are known, it is required to establish the emission algorithm which decides whether a particle is being emitted from the compound nucleus. This is done \cite{Chaudhuri} by first calculating the ratio $x=\tau / \tau_{tot}$ where $\tau_{tot}= \hbar / \Gamma_{tot}$, $\Gamma_{tot}=\sum_{\nu}\Gamma_{\nu}$ and $\nu =n,p,d,t,He^3,\alpha,\gamma$ or fission and then performing Monte-Carlo sampling from a uniformly distributed set of random numbers. In the case that a particle is emitted, the type of the emitted particle is next decided by a Monte Carlo selection with the
weights $\Gamma_{\nu}/\Gamma_{tot}$ (partial widths). The energy of the emitted particle is then obtained by another Monte Carlo sampling of its energy spectrum. The energy, mass and charge of the nucleus is adjusted after each emission and the entire procedure is repeated until the resulting products are unable to undergo further decay. This procedure is followed for each of the primary fragment produced at a fixed temperature and then repeated over a large
ensemble and the observables are calculated from the ensemble averages. The number and type of particles emitted and the final decay product in each event is registered and are taken into account properly keeping in mind the overall charge and baryon number conservation. This is the third and final stage of the calculation. The same calculation is repeated for each set of fragments produced after multifragmentation at different temperatures.

\section{Temperature profile in projectile fragmentation}

Initially with the increase of the projectile beam energy, the temperature of the projectile spectator also increases. But above a certain energy of the projectile beam the temperature of the projectile spectator will not increase. This is known as limiting fragmentation. This projectile fragmentation model is valid in the limiting fragmentation region. The main reasons behind the excitation of projectile spectator are highly non-spherical shape and migration of some nucleons from participant to projectile spectator.   Without a calculation at a more fundamental level it is not possible to calculate the excitation.  At this stage we do not deal with excitation energy as such and characterize the system by a temperature $T$. We plan to work on this in future. Though the temperature is independent of incident energy of the projectile, it depends upon the impact parameter.\\

To get the impact parameter dependent temperature profile i.e. $T=T(b)$ two types of parametrization comes to mind. The simplest case is that the temperature directly depends upon the impact parameter i.e. $T(b)=C_0+C_1b+C_2b^2+...$. In the another parametrization the temperature depends on the wound that the projectile suffers in the collision i.e. $1.0-A_s(b)/A_0$, so in this case $T(b)=D_0+D_1(A_s(b)/A_0)+D_2(A_s(b)/A_0)^2+...$. After calculating different observables of projectile fragmentation by using temperature profiles and comparing the theoretical results with experimental data, it is observed that linear parameterizations are enough i.e. $C_2$, $C_3$... or $D_2$, $D_3$... are negligible. Since $T(b)=D_0+D_1(A_s(b)/A_0)$ is physically more acceptable than $T(b)=C_0+C_1b$, we finally choose this temperature profile. We fix the values of $D_0$ and $D_1$ by comparing our model results with experimental data of mass distribution and multiplicity of intermediate mass fragments of different target-projectile combinations. The comparison led to the values $D_0$=7.5 and $D_1$=-4.5 which are used for subsequent calculations.\\

\begin{figure}[t]
\begin{center}
\includegraphics[height=2.65in,width=4.5in,]{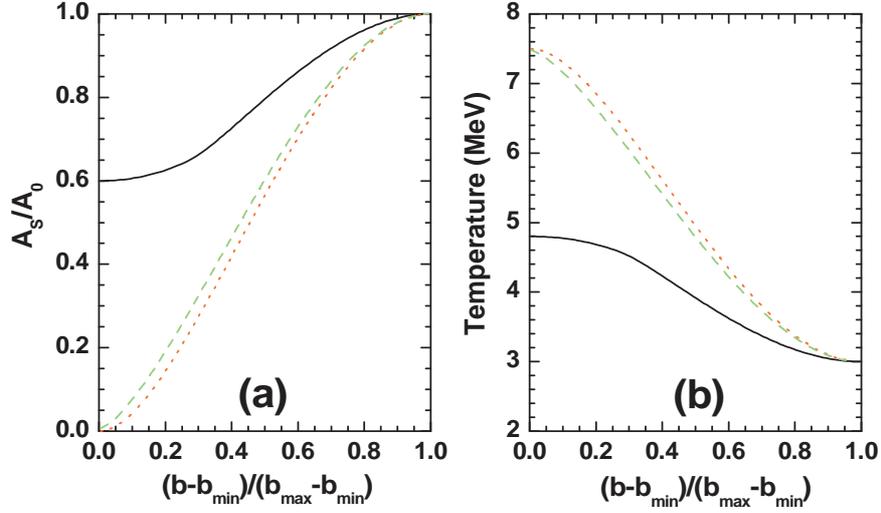}
\label{fig1}
\caption{ Variation of (a) $A_s/A_0$ and (b) temperature with normalized impact parameter $(b-b_{min})/(b_{max}-b_{min})$ for $^{58}$Ni on $^9$Be (solid line), $^{58}$Ni on $^{181}$Ta (dotted line) and $^{124}$Sn on $^{119}$Sn (dashed line) reactions.}
\end{center}
\end{figure}
For three different nuclear reactions $^{58}$Ni on $^9$Be, $^{58}$Ni on $^{181}$Ta and $^{124}$Sn on $^{119}$Sn, the variation of the quantity $A_s/A_0$ obtained after abrasion stage with normalized impact parameter $(b-b_{min})/(b_{max}-b_{min})$ is shown in Fig. 1.a where as Fig. 1.b represents the freeze-out temperature profile of these three reactions calculated from the formula $T(b)=7.5-4.5(A_s(b)/A_0)$. The specification that $T(b)=D_0+D_1(A_s(b)/A_0)$ has profound consequences.  This means the temperature profile $T(b/b_{max})$ of $^{124}$Sn on $^{119}$Sn is very different from that of $^{58}$Ni on $^9$Be. In the first case $A_s(b)/A_0$ is nearly zero for $b=b_{min}$=0 whereas in the latter case $A_s(b)/A_0$ is $\approx 0.6$ for $b=b_{min}$=0. Even more remarkable feature is that the temperature profile of $^{58}$Ni on $^9$Be is so different from the temperature profile of $^{58}$Ni on $^{181}$Ta. In the latter case $b_{min}=R_{Ta}-R_{Ni}$ and beyond $b_{min}$, $A_s(b)/A_0$ grows from zero to 1 for $b_{max}$. This is very similar to the temperature profile of $^{124}$Sn on $^{119}$Sn.\\

\begin{figure}[b]
\begin{center}
\includegraphics[height=2.65in,width=4.5in,]{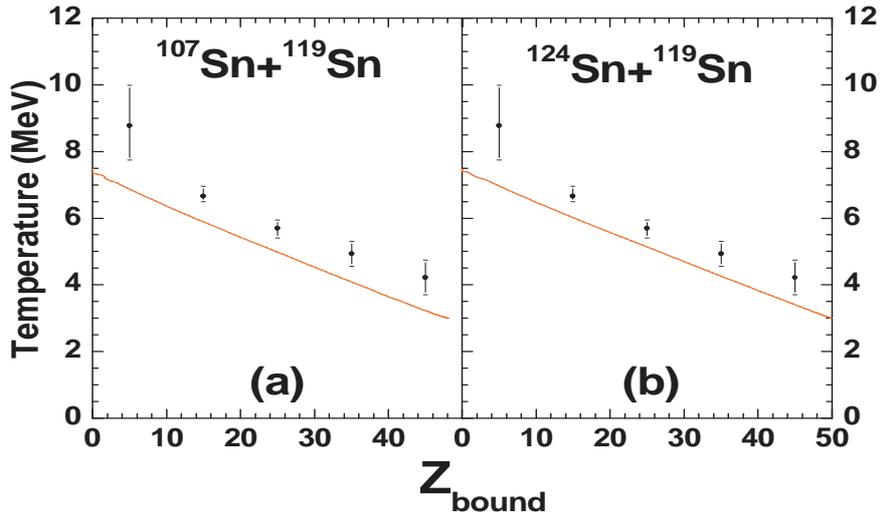}
\label{fig2}
\caption{ Comparison of theoretically used temperature profile calculated by the formula  $T(b)=7.5-4.5(A_s(b)/A_0)$ (solid lines) with that deduced by Albergo formula from experimental data \cite{Ogul} (circles with error bars) for (a) $^{107}$Sn on $^{119}$Sn and (b) $^{124}$Sn on $^{119}$Sn.}
\end{center}
\end{figure}
In Fig.2 the temperatures calculated from the model is plotted with $Z_{bound}$ (=$Z_s$ minus charges of all composites with charge $Z=1$) and compared with experimentally measured temperatures (by Albergo formula \cite{Albergo}) for two different projectile fragmentation reactions $^{107}$Sn on $^{119}$Sn and $^{124}$Sn on $^{119}$Sn. The experiments are done by ALADIN collaboration in GSI at 600A MeV \cite{Ogul}.

\section{Results}

The projectile fragmentation model is used to calculate the basic observables of projectile fragmentation like the average number of intermediate mass fragments ($M_{IMF}$), the average size of the largest cluster and their variation with bound charge ($Z_{bound}$), differential charge distribution, total charge distribution, isotopic distribution for different nuclear reactions at intermediate energies with different projectile target combinations.

\subsection{$M_{IMF}$ variation with $Z_{bound}$}
\begin{figure}[h]
\begin{center}
\includegraphics[height=2.65in,width=4.5in]{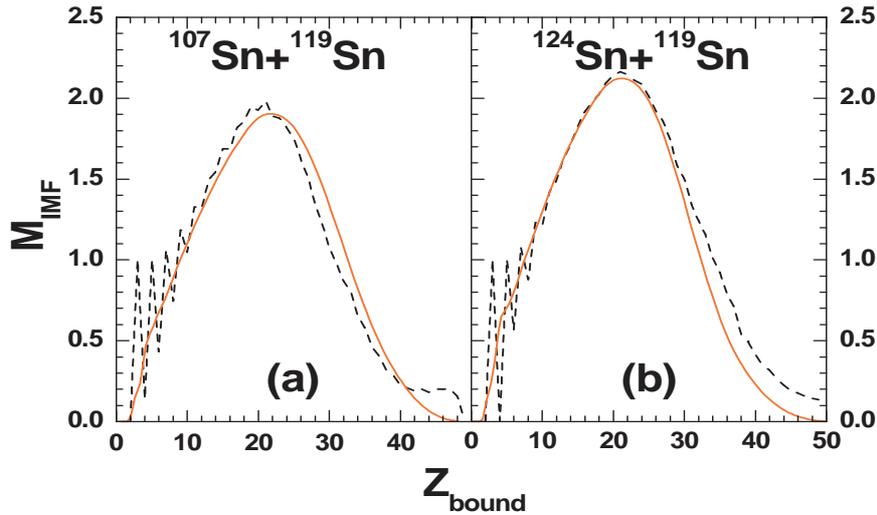}
\label{fig3}
\caption{ Mean multiplicity of intermediate-mass fragments $M_{IMF}$, as a function of $Z_{bound}$ for (a) $^{107}$Sn on $^{119}$Sn and (b) $^{124}$Sn on $^{119}$Sn reaction obtained from projectile fragmentation model (solid lines). The experimental results are shown by the dashed lines. }
\end{center}
\end{figure}
The variation of the average number of intermediate mass fragments $M_{IMF}$ ($3{\le}Z{\le}20$) with $Z_{bound}$ for $^{107}$Sn on $^{119}$Sn and $^{124}$Sn on $^{119}$Sn reactions is shown in Fig.3. The theoretical calculation reproduces the average trend of the experimental data very well. At small impact parameters, the size of the projectile spectator (also $Z_{bound}$) is small and the temperature of the dissociating system is very high. Therefore the PLF will break into fragments of small charges (mainly $Z=1, 2$). Therefore the IMF production is less. But at mid-central collisions PLF's are larger in size and the temperature is smaller compared  to the  previous case, therefore larger number of IMF's are produced. With further increase of impact parameter, though the PLF size (also $Z_{bound}$) increases,  the temperature is low, hence breaking of dissociating system is very less (large fragment remains) and therefore IMF production is less.
\subsection{Differential charge distribution}
\begin{figure}[h]
\begin{center}
\includegraphics[height=2.65in,width=4.5in]{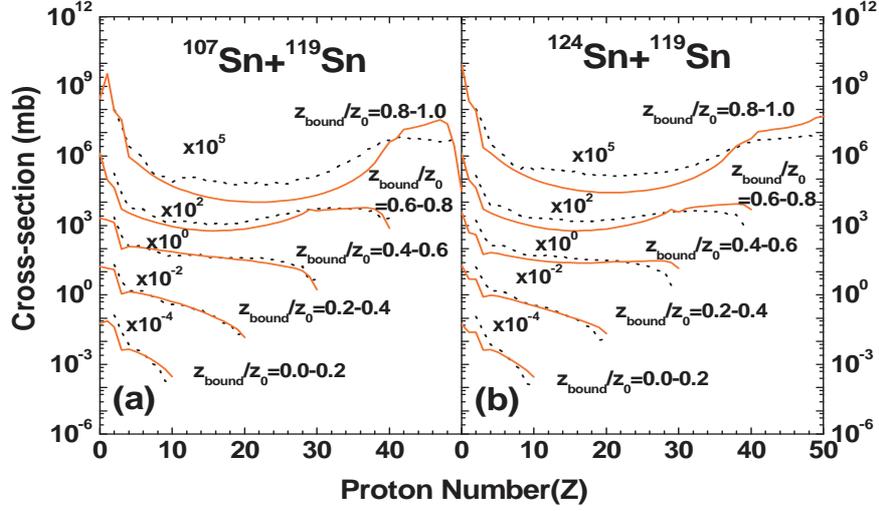}
\label{fig4}
\caption{ Theoretical differential charge cross-section distribution (solid lines) for (a) $^{107}$Sn on $^{119}$Sn and (b) $^{124}$Sn on $^{119}$Sn reaction compared with the experimental data (dashed lines).}
\end{center}
\end{figure}
The differential charge distributions for different intervals of $Z_{bound}/Z_0$ are calculated by the projectile fragmentation model for $^{119}$Sn and $^{124}$Sn on $^{119}$Sn reactions and compared with experimental data \cite{Ogul}. This is shown in Fig.4. For the sake of clarity the distributions are normalized with different multiplicative factors. At peripheral collisions (i.e. $0.8{\le}Z_{bound}/Z_0{\le}1.0$)  due to small temperature of PLF, it breaks into one large fragment and small number of light fragments, hence the charge distribution shows $U$ type nature. But with the decrease of impact parameter the temperature increases, the PLF breaks into larger number of fragments and the charge distributions become steeper. The features of the data are nicely reproduced by the model.

\subsection{Size of largest cluster and its variation with $Z_{bound}$}
\begin{figure}[b]
\begin{center}
\includegraphics[height=2.65in,width=4.5in]{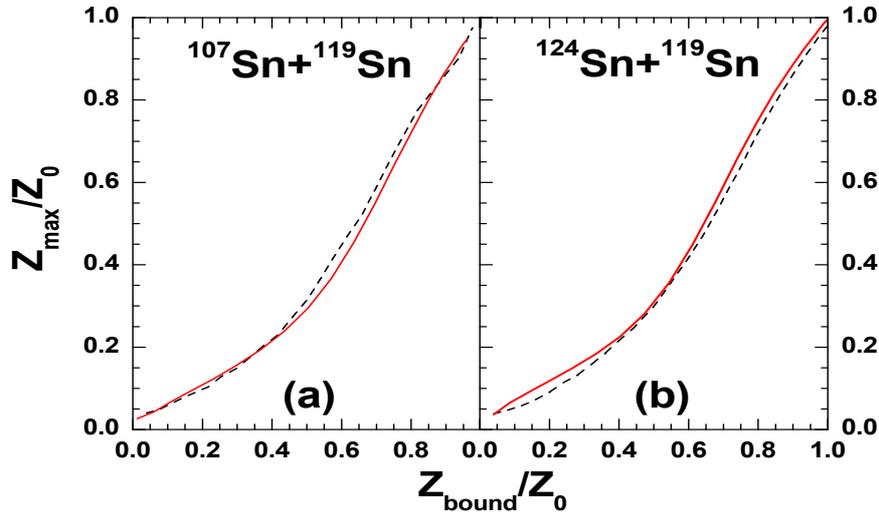}
\label{fig5}
\caption{ $Z_{max}/Z_0$ as a function of $Z_{bound}/Z_0$ for (a) $^{107}$Sn on $^{119}$Sn and (b) $^{124}$Sn on $^{119}$Sn reaction obtained from projectile fragmentation model (solid lines). The experimental results are shown by the dashed lines. }
\end{center}
\end{figure}
Average size of the largest cluster produced at different $Z_{bound}$ values is calculated in the framework of projectile fragmentation model for $^{119}$Sn and $^{124}$Sn on $^{119}$Sn reactions.  In Fig.5 the variation of $Z_{max}/Z_0$ ($Z_{max}$ is the average number of proton content in the largest cluster) with $Z_{bound}/Z_0$ obtained from theoretical calculation and experimental result are shown. Very nice agreement with experimental data is observed.

\subsection{Total charge distribution}

The total charge distributions of different experiments ($^{58}$Ni on $^{9}$Be and $^{58}$Ni on $^{181}$Ta at 140 MeV/nucleon done at MSU \cite{Mocko}, $^{129}$Xe on $^{29}$Al at 790 MeV/nucleon \cite{Reinhold} and $^{136}$Xe on $^{208}$Pb at 1 GeV/nucleon \cite{Henzlova} both performed at GSI) are theoretically calculated from the projectile fragmentation model by using same temperature profile. This is shown Fig.6. In theoretical calculation cross-section of all fragments ranging from light nucleon to original projectile are calculated separately, but in Fig.6 the cross-section of the fragments for which experimental data is available is only shown. It is observed that, though the projectile beam energies in experiments are widely different, the same temperature profile can explain all the data pretty well.
\begin{figure}[h]
\begin{center}
\includegraphics[width=3.4in,height=3.4in,clip]{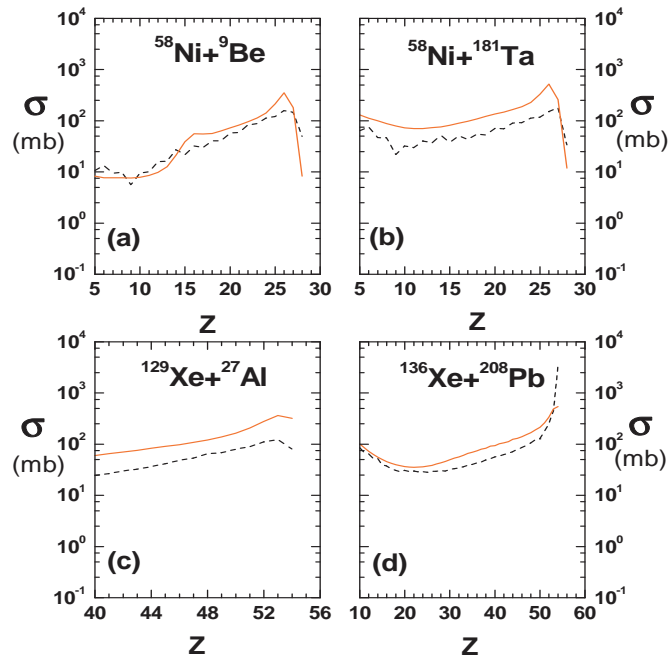}
\label{fig7}
\caption{ Theoretical total charge distribution (solid lines) for (a) $^{58}$Ni on $^{9}$Be, (b) $^{58}$Ni on $^{181}$Ta, (c) $^{129}$Xe on $^{29}$Al and (d) $^{136}$Xe on $^{208}$Pb reaction compared with experimental data (dashed lines).}
\end{center}
\end{figure}
\subsection{Isotopic distribution}
\begin{figure}[h]
\begin{center}
\includegraphics[width=4.6in,height=3.2in,clip]{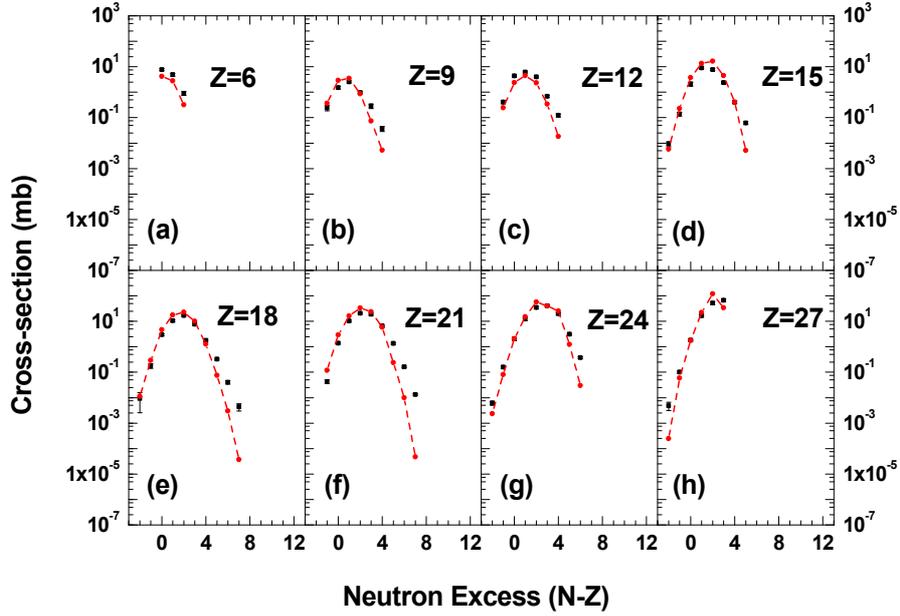}
\label{fig7}
\caption{ Theoretical isotopic cross-section distribution (circles joined by dashed lines) for $^{58}$Ni on $^{9}$Be reaction compared with experimental data \cite{Mocko} (squares with error bars).}
\end{center}
\end{figure}
For $^{58}$Ni on $^{9}$Be reaction at 140 MeV/nucleon \cite{Mocko}, the isotopic distributions are theoretically calculated and compared with experimental values. This is shown in Fig.7. Nice agreement between theoretical result and experimental data is obtained.
\section{Summary and Conclusion}
A model for projectile fragmentation is developed which is grounded on the traditional concepts of heavy-ion reaction (abrasion) as well as the model of multifragmentation (Canonical thermodynamical model) and secondary decay. This model is in general applicable and implementable in the limiting fragmentation region. An impact parameter dependent temperature profile is introduced in the model for projectile fragmentation which could successfully explain experimental data of different target projectile combinations of widely varying projectile energy. The observables which are calculated and compared to data included charge distribution, isotopic distribution intermediate mass fragment multiplicity and average size of largest cluster. \\
While we have reasonable agreements with many data considered here, it is desirable to push the model for further improvements. The goal will be to find the size and excitation of the initial projectile spectator from microscopic BUU calculations. We plan to work on this.
\section{Acknowledgments}
The authors are thankful to Prof. Wolfgang Trautmann, Dr. M. Mocko and Dr. Karl-Heinz Schmidt for access to experimental data.

\end{document}